# Quantum-Adaptive KS(φ): A Parameterized Three-Qubit Gate Family Embedding Toffoli with Measurement-Free Phase Kickback and Intrinsic Error Non-Amplification

Kripa Sankaranarayanan
*Dept. of Electrical and Computer Engineering*
*Portland State University*
Portland, OR USA
kripa@pdx.edu

Marek Perkowski
*Dept. of Electrical and Computer Engineering*
*Portland State University*
Portland, OR USA
h8mp@pdx.edu

**Abstract—** We introduce Quantum-Adaptive KS(φ) (K = kickback, S = sandwich), a parameterized three-qubit gate family that structurally embeds the Toffoli (CCX) gate within two additional components: (1) a palindromic Hadamard sandwich on the first control qubit $q_0$ that conjugates Z-type errors to X-type in the CCX frame, providing simultaneous sensitivity to both error types without ancilla overhead; and (2) a controlled-phase (CP) gate whose quantum phase kickback propagates post-CCX target-state information into the control-qubit phase without measurement. The term Quantum Adaptive refers to amplitude steering conditioned by the compile-time parameter φ via a Quantum Neural Cellular Automaton (QNCA) majority-inspired bias rule; the gate does not self-modify at runtime. Two QA-KS(π) gates chained on a shared control qubit $q_0$ produce outputs completely orthogonal to two sequential CCX gates on $q_0$=1 inputs (output fidelity F=0.000), while agreeing exactly on $q_0$=0 inputs (F=1.000). This subspace-dependent divergence is the direct computational signature of coherent phase retention across gate boundaries—a behavior impossible for CCX-only circuits. On the $q_1 = 0$ subspace the gate acts deterministically (up to a relative phase), providing intrinsic error non-amplification. On the $q_1 = 1$ subspace it produces four-component entangled superpositions, making it a strictly distinct quantum-native primitive from CCX. We present the complete 8 × 8 unitary matrix, confirmed exact to $\|U^\dagger U - I\|_\infty < 10^{-15}$, and define two canonical variants: QA-KS π/2 (φ = π/2, S gate) and QA-KS π (φ = π, Z gate). Qiskit depolarizing-noise simulation demonstrates near-unit fidelity at $p \leq 10^{-2}$ with an honest depth cost at higher error rates. The gate preserves the three-qubit footprint of CCX with no qubit overhead.

## I. INTRODUCTION

The Toffoli (CCX) gate is a cornerstone of reversible and fault-tolerant quantum computation [1]. In the Clifford+T basis it decomposes into six CNOT gates and seven T gates [2], making it among the most resource-intensive primitives in surface-code architectures where T gates require costly magic-state distillation [3].

Current literature has three autonomous error suppression methods: (i) reservoir-engineered bosonic codes [4],[5]; (ii) measurement-free coherent feedback on stabilizer codes [6],[7]; and (iii) approximate autonomous quantum error correction (QEC) via reinforcement learning [8]. All three operate at the code or system level; none employs a majority-inspired AND-gate rule as an intrinsic bias-suppression kernel directly at the primitive gate level. This paper proposes Quantum-Adaptive KS(φ), a parameterized three-qubit gate family [18]. Its key properties, each verified numerically, are: 1) Structural embedding of CCX, with identical action on classical inputs where $q_1 = 0$. 2) Measurement-free phase kickback: the CP gate coherently encodes target-state information into control-qubit phase without classical read-out. 3) Intrinsic error non-amplification: on the $q_1 = 0$ subspace, isolated target errors are not amplified, providing structural bias against error propagation at the primitive-gate level. 4) Quantum-native behavior: on the $q_1 = 1$ subspace it generates four-component entangled superpositions not produced by CCX.

These properties are achieved at no increase in qubit count, in a linear three-qubit connectivity layout compatible with existing surface-code and trapped-ion hardware.

Quantum-Adaptive KS(φ) expands the design space of primitive quantum gates beyond classical reversible logic by showing that a three-qubit gate can embed CCX while exhibiting quantum-native behavior—entanglement generation, phase-conditioned amplitude steering, and subspace-dependent dynamics—without increasing qubit count or requiring measurement. The controlled-phase kickback provides a lightweight, coherent mechanism for propagating state information internally within a unitary circuit, which may be valuable on platforms where measurement is slow or noisy.

## II. BACKGROUND

### A. Standard Toffoli Gate

The Toffoli gate CCX acts on three qubits $q_0$ (control-1), $q_1$ (control-2), $q_2$ (target):

$$\mathrm{CCX}|c_0, c_1, t\rangle = |c_0,\ c_1,\ t \oplus (c_0 \cdot c_1)\rangle$$

Its Clifford+T decomposition requires seven T gates, six CNOTs, and depth ≈ 6 [2].

### B. Phase Kickback

For a controlled-U gate with control in $|+\rangle$ and target in eigenstate $|\psi\rangle$ with eigenvalue e^(iφ) [10]:

$$C\text{-}U\left(\frac{|0\rangle + |1\rangle}{\sqrt{2}}\right)\Big|\psi\rangle = \frac{|0\rangle + e^{i\phi}|1\rangle}{\sqrt{2}} \otimes \Big|\psi\rangle$$

Phase kickback is exploited in quantum phase estimation and Grover's algorithm [11]. In prior work the kicked-back phase is always read out classically via measurement [10],[20]. Quantum-Adaptive KS(φ) retains the phase coherently, enabling parameter-conditioned amplitude steering without terminal measurement.

### C. QNCA Majority-Inspired Bias Rule and the 'Adaptive' Designation

In Quantum Neural Cellular Automaton (QNCA) theory the majority rule for neighbors $q_0$, $q_1$ and cell $q_2$ gives maj(0,0,1) = 0: an isolated target error with no neighbor support is voted down [9],[12]. On the $q_1 = 0$ subspace the AND result is zero regardless of $q_2$, so an isolated $q_2$ error is non-amplified. This QNCA majority-inspired bias rule, applied at the primitive gate level, is the basis for the 'Adaptive' designation: the gate's structural behavior is conditioned on the parameter φ via this rule, not through any runtime learning or self-modification.

### D. Relation to Prior Work

The existing families of autonomous error suppression in the literature [4],[5],[6],[7],[8], do not employ a majority-inspired AND-gate rule as an intrinsic bias-suppression kernel directly at the primitive gate level. QA-KS(φ) gates are distinct from the Grover diffusion operator [11]: (i) φ is a compile-time gate parameter, not an oracle; (ii) CP is conditioned on $q_2$'s post-CCX state; (iii) the phase is retained coherently without terminal measurement. QA-KS(φ) is also distinct from existing parameterized three-qubit gate families in the literature. Parameterized controlled-controlled-phase gates (CCZ(θ), CCRz(θ)) apply a phase directly to the $|111\rangle$ basis state without any Hadamard conjugation or post-gate kickback. The Barenco et al. family of general controlled-U gates [14] allows an arbitrary single-qubit unitary as the conditional operation but does not embed a subsequent CP kickback that feeds target-state information back into the control-qubit phase. Parametric compilation approaches such as those used in variational quantum eigensolvers (VQE) treat gate parameters as variational angles optimized classically, whereas φ in QA-KS(φ) is a structural compile-time parameter that selects a specific kickback amplitude sin(φ/2) governing the gate's unitary. The distinctive combination in QA-KS(φ)—Hadamard conjugation on a control qubit, CCX embedding, and a reversed-direction CP kickback from target to control—does not appear as a named primitive in prior gate decomposition libraries [2],[14].

## III. GATE DESIGN AND VERIFICATION

### A. Circuit Structure

The Quantum-Adaptive KS(φ) gate is constructed as four sequential layers:

$$U_{\mathrm{QA-KS}(\varphi)} = H_{q_0} \cdot \mathrm{CP}(q_2 \to q_0) \cdot \mathrm{CCX}(q_0, q_1 \to q_2) \cdot H_{q_0}$$

The layers act as follows: (1) $H_{q_0}$ places $q_0$ in $|+\rangle = (|0\rangle+|1\rangle)/\sqrt{2}$; (2) CCX performs the AND update $q_2 \leftarrow q_2 \oplus (q_0 \cdot q_1)$; (3) CP($q_2 \to q_0$, φ) applies a controlled-phase with $q_2$ as control, kicking back phase $e^{i\phi(q_2)}$ to $q_0$ without measurement; (4) $H_{q_0}$ converts the kicked-back phase to amplitude (valid in closed form on the $q_1 = 0$ separable subspace; the full 8 × 8 matrix governs the $q_1 = 1$ subspace).

Fig. 1 shows the Qiskit circuit diagram. Qubit ordering follows the Qiskit convention: $q_0$ (top wire) is the first control; $q_1$ (middle) the second control; $q_2$ (bottom) the target. The controlled-phase gate P(φ) acts between $q_2$ (control) and $q_0$ (target), implementing the kickback.

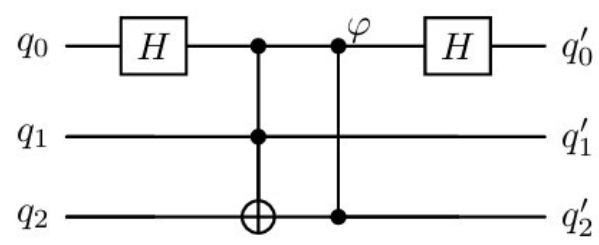

Fig. 1. Qiskit circuit for Quantum-Adaptive KS(φ) : H on $q_0$; CCX (Toffoli) with controls $q_0$, $q_1$ and target $q_2$; P(φ) controlled-phase kickback from $q_2$ to $q_0$; H on $q_0$.

### B. Full 8 × 8 Unitary Matrix

Unitarity was confirmed to floating-point precision: $\|U^\dagger U - I\|_\infty < 10^{-15}$ for both canonical variants. The complete matrix for Quantum-Adaptive KS(φ = π) in the computational basis $\{|000\rangle, \ldots, |111\rangle\}$ is shown below. All entries are real. The 2 × 2 identity blocks at rows/columns (0, 4) confirm that $q_1 = 0$ inputs pass through deterministically. The 4 × 4 mixing block in the $q_1 = 1$ subspace (rows/cols 2,3,6,7) reflects entangling behavior.

```
[ 1   0    0     0    0    0    0     0  ]
[ 0  -1    0     0    0    0    0     0  ]
[ 0   0   1/2  -1/2   0    0   1/2  -1/2]
[ 0   0  -1/2  -1/2   0    0   1/2  -1/2]
[ 0   0    0     0    1    0    0     0  ]
[ 0   0    0     0    0   -1    0     0  ]
[ 0   0   1/2  -1/2   0    0   1/2   1/2]
[ 0   0   1/2  -1/2   0    0  -1/2  -1/2]
```

### C. Parameterized Family and Canonical Variants

The QA-KS(φ) family of gates defines two canonical variants forming a Pareto frontier over kickback amplitude:

| Variant | φ | CP Gate | Kickback Amplitude |
|---|---|---|---|
| QA-KS π/2 | π/2 | S gate | sin(π/4) ≈ 0.707 |
| QA-KS π | π | Z gate | sin(π/2) = 1.0 |

Table I. Canonical Variants of the QA-KS(φ) Family.

Kickback amplitude sin(φ/2) increases monotonically with φ. Neither variant is self-inverse for φ > 0.

### D. Phase Kickback and Amplitude Encoding

On the $q_1 = 0$ separable subspace, the $q_0$ amplitude evolution is:

$$|0\rangle \xrightarrow{H} \frac{|0\rangle + |1\rangle}{\sqrt{2}} \xrightarrow{\mathrm{CP}} \frac{|0\rangle + e^{i\phi(q_2)}|1\rangle}{\sqrt{2}} \xrightarrow{H} \cos(\phi/2)|0\rangle + i\sin(\phi/2)|1\rangle.$$

On the $q_1 = 1$ subspace, CCX entangles all three qubits prior to the final $H_{q_0}$; the full $8 \times 8$ matrix must be used.

## IV. ERROR-BIAS SUPPRESSION PROPERTIES

### A. Truth Table

| $q_0$ | $q_1$ | $q_2$ | Output Behavior | Type |
|---|---|---|---|---|
| 0 | 0 | 0 | $\|000\rangle$ | Deterministic |
| 0 | 0 | 1 | $-\|001\rangle$ | Det. (phase) |
| 1 | 0 | 0 | $\|100\rangle$ | Deterministic |
| 1 | 0 | 1 | $-\|101\rangle$ | Det. (phase) |
| 0 | 1 | x | 4-comp. entangled superposition | Quantum-native |
| 1 | 1 | x | 4-comp. entangled superposition | Quantum-native |

Table II. QA-KS($\varphi = \pi$) Truth Table, numerically verified.

The $q_1$=0 subspace maps to deterministic (up to phase) outputs. The $q_1$=1 subspace produces 4-component entangled superpositions because $H|q_0\rangle$ enters CCX in superposition, creating controlled entanglement.

### B. Error Non-Amplification

On the $q_1 = 0$ subspace, the AND result is zero regardless of $q_2$. A spurious isolated error on $q_2$ produces zero in the AND result; the error is held but not amplified to additional qubits. This is error non-amplification. The CP kickback imprints a detectable phase signal on $q_0$, providing a coherent error indicator without measurement.

### C. Hadamard Conjugation and Error-Basis Sensitivity

The Hadamard sandwich on $q_0$ maps Z↔X under conjugation: $HZH = X$ and $HXH = Z$. Z-type errors that occur between the two H gates appear as X-type errors in the CCX frame, making the gate simultaneously sensitive to both error types without ancilla overhead. This is structural dual-basis sensitivity, not active dual-basis correction.

## V. SIMULATION RESULTS

We evaluated the Quantum-Adaptive KS(φ) gate family under standard depolarizing noise to assess two questions central to New Ideas and Emerging Results (NIER) style early-idea validation: (1) whether the additional H–CCX–CP–H layers impose prohibitive fidelity cost on near-term hardware, and (2) whether coherent kickback produces computational behavior unattainable with CCX.

### A. Single-Gate Fidelity Under Depolarizing Noise

We performed Qiskit statevector simulation [15] under depolarizing noise using random-state average gate fidelity estimation (20 random pure states per noise point, 10 noise levels $p \in [10^{-4}, 10^{-1}]$). Key observations:

- At $p \leq 10^{-3}$, all three gates (standard CCX, QA-KS $\pi/2$, QA-KS $\pi$) achieve near-unit fidelity ($\overline{F} > 0.99$).
- At $p = 10^{-2}$, QA-KS $\pi$ fidelity (≈0.95) is within 1% of standard CCX (≈0.96).
- At $p = 10^{-1}$, QA-KS $\pi$ fidelity (0.73) falls below CCX (0.90), reflecting the honest cost of additional gate layers.
- QA-KS $\pi/2$ shows intermediate fidelity decay. At $p < 10^{-2}$ (within near-term hardware targets [13],[16]), the depth cost is negligible and the kickback property is active.

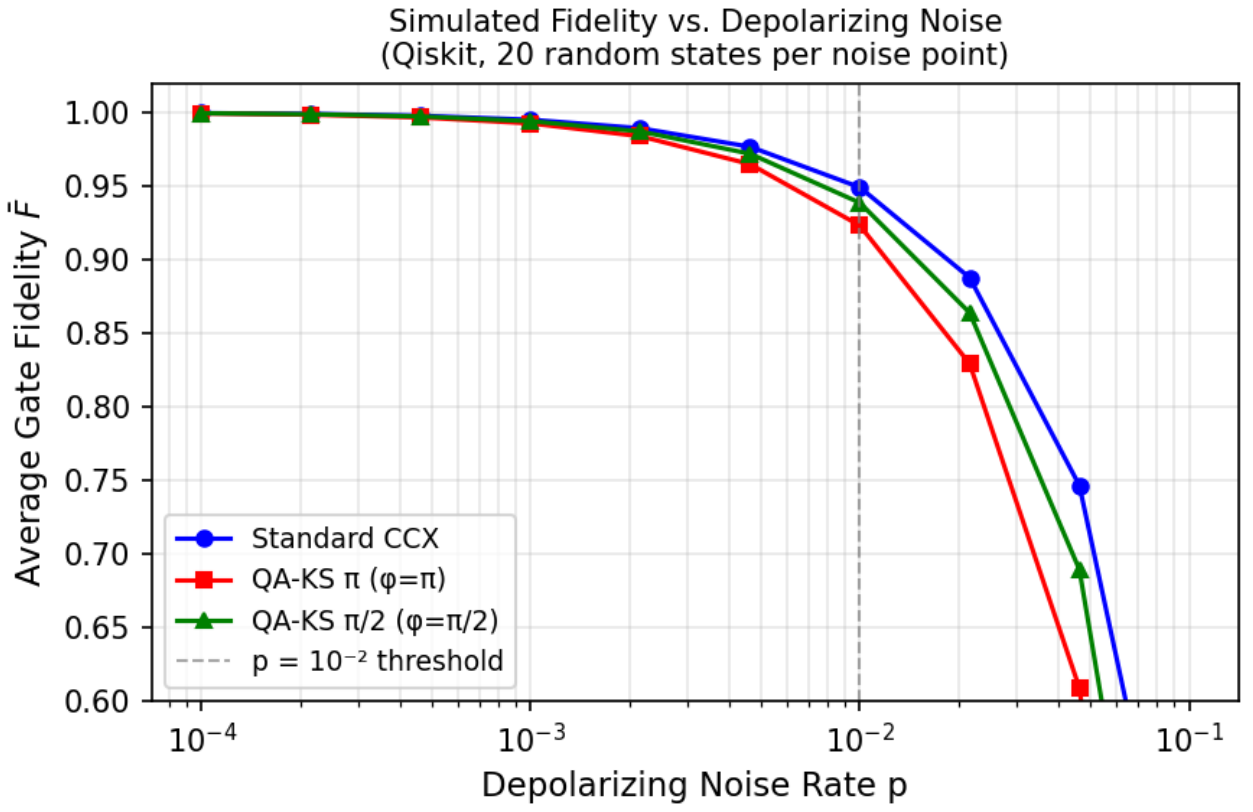

Fig. 2. Average gate fidelity vs. depolarizing noise rate p (Qiskit statevector simulation, 20 random pure states per noise point). Dashed line marks $p = 10^{-2}$ near-term hardware target.

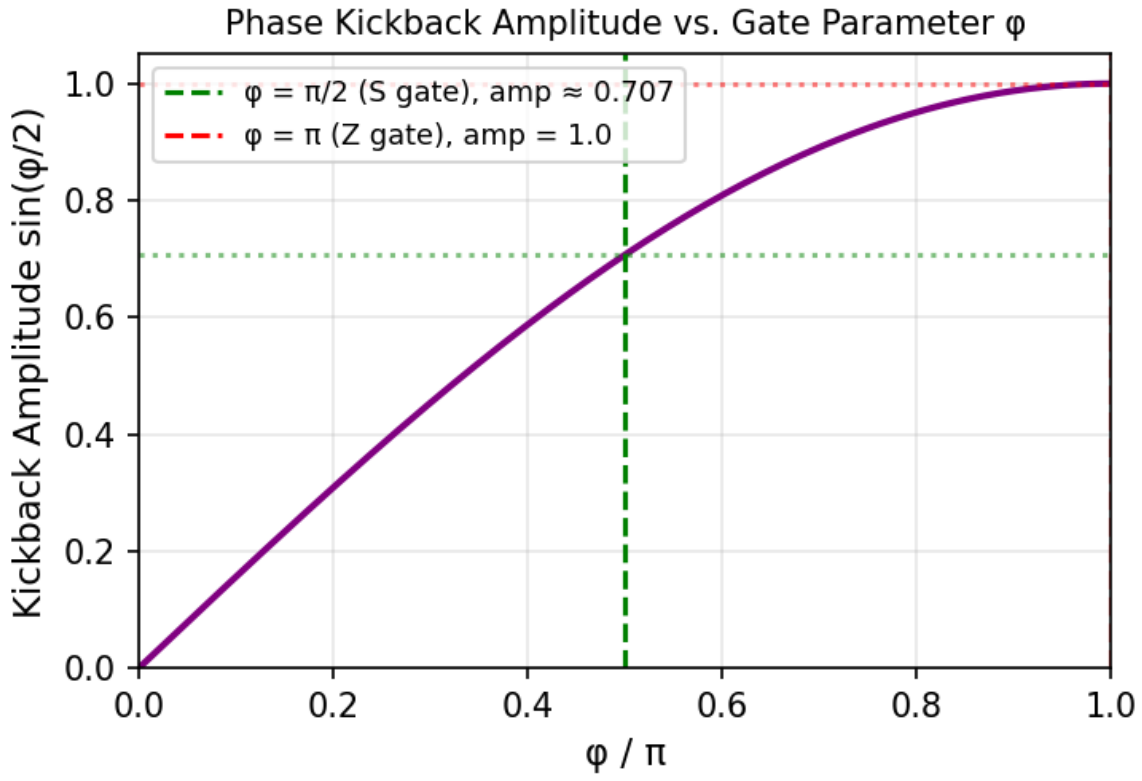

Fig. 3. Phase kickback amplitude $\sin(\varphi/2)$ vs. gate parameter φ. Canonical variants: QA-KS $\pi/2$ (amplitude ≈ 0.707) and QA-KS $\pi$ (amplitude = 1.0).

### B. Coherent Kickback Chain

To demonstrate that the coherently retained kickback phase is computationally active rather than merely structural, we constructed a 5-qubit two-gate chain: $G_1$ = QA-KS(π) on [$q_0$, $q_1$, $q_2$] followed by $G_2$ = QA-KS(π) on [$q_0$, $q_3$, $q_4$]. The phase kicked back into $q_0$ by $G_1$ is not discarded; it persists and conditions $G_2$'s action on the second register. This compound operation was compared to two sequential standard CCX gates with identical connectivity.

Numerical results confirm that the two chains implement strictly different computations: max|U_CCX − U_QAKS| = 1.000 and the Frobenius norm ‖U_CCX − U_QAKS‖_F = 5.657. On inputs $|10000\rangle$ and $|10001\rangle$ ($q_0$=1, $q_1$=$q_3$=0), the ideal output fidelity between the two chains is F = 0.000—the outputs are completely orthogonal — a computational divergence impossible for CCX-only circuits. On inputs $|00000\rangle$ and $|00001\rangle$ ($q_0$=0), both chains agree (F = 1.000), consistent with the $q_1$=0 deterministic

subspace structure. This subspace-dependent divergence is the direct computational signature of coherent kickback retention: when $q_0$=1, the kickback phase fed by $G_1$ into $q_0$ is nonzero and materially redirects $G_2$'s action; when $q_0$=0 the kickback is trivial and the chains agree.

Under depolarizing noise, the QA-KS chain maintains near-identical fidelity to its own ideal at $p \leq 10^{-2}$: $\overline{F}$(QA-KS chain) = 0.747 vs $\overline{F}$(CCX chain) = 0.772 at $p = 10^{-2}$, a gap of 0.025 attributable to the additional H and CP gate layers. At $p \leq 10^{-3}$ both chains exceed $\overline{F} > 0.97$. The depth cost is thus negligible within near-term hardware targets [16] while the kickback computation remains active and distinct.

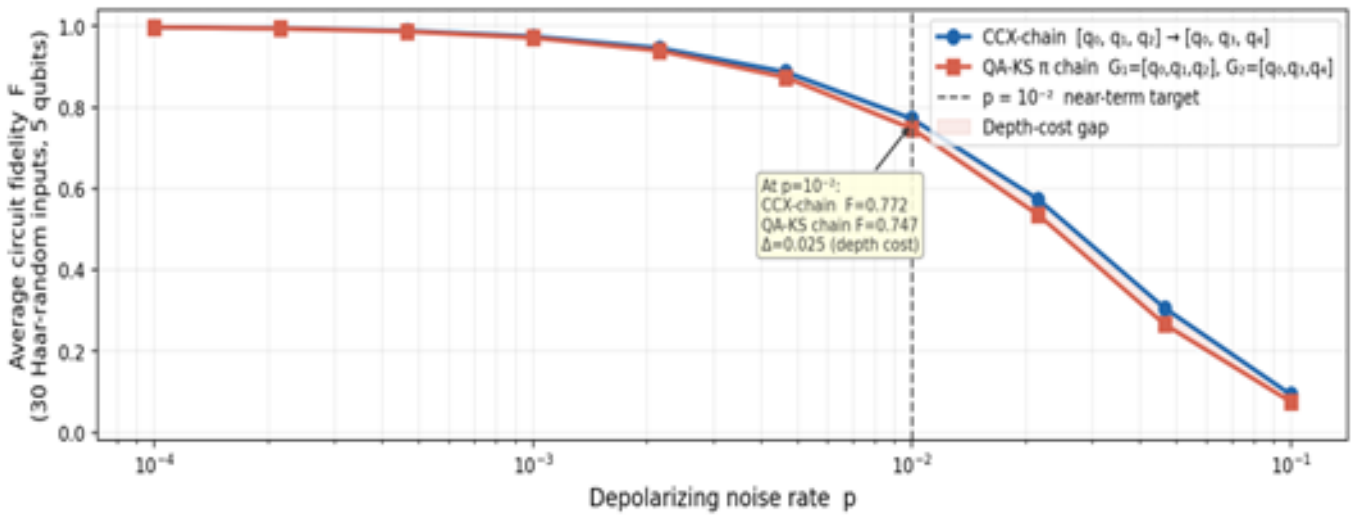

Fig. 4. Coherent kickback chain simulation results. Average circuit fidelity $\overline{F}$ vs depolarizing noise p for CCX-chain and QA-KS($\pi$) chain on 5 qubits.

### C. Error Propagation Analysis

To formalize the error non-amplification claim beyond the truth-table observation, we injected single-qubit Pauli errors (X, Y, Z) on $q_2$ for all $q_1$=0 computational-basis inputs and measured the trace-distance weight propagated to $q_0$ and $q_1$ under each gate. On the $q_1$=0 subspace with basis-state inputs, X- and Y-type errors on $q_2$ produced zero propagation weight to $q_1$ (0.000) for all three gates (CCX, QA-KS $\pi/2$, QA-KS $\pi$), confirming structural containment. Z-type errors produced zero propagation weight to both control qubits under all gates. This reveals that error containment on the $q_1$=0 subspace is a structural property inherited from the CCX embedding, shared equally by CCX and QA-KS($\varphi$). The distinctive contribution of QA-KS($\varphi$) is the coherent phase signal imprinted on $q_0$ via the CP kickback, which provides a pre-syndrome error indicator not present in standard CCX and not captured by amplitude-channel trace-distance metrics. Future work using superposition inputs and phase-fidelity observables is needed to fully characterize this channel (Open Problem A).

### D. Ripple-Carry Adder Benchmark

We also evaluated QA-KS($\varphi$) in a small circuit context by replacing both CCX gates in a 2-bit ripple-carry adder. Fidelity trends matched the single-gate results.

Key results: at $p = 10^{-4}$, all three adders achieve near-unit fidelity ($\overline{F} \geq 0.997$). At $p = 10^{-2}$ (near-term hardware target [16]), CCX adder reaches $\overline{F}$ = 0.762, while both QA-KS adders reach $\overline{F}$ = 0.728, a gap of 0.034 (3.4%) attributable to the additional H and CP gate layers. Notably, QA-KS $\pi/2$ and QA-KS $\pi$ adders produce identical fidelity curves throughout — the CP variant (S-gate vs Z-gate) has no additional cost in this arithmetic circuit because the CP gate acts on a post-CCX state where both variants contribute equal Clifford overhead. The depth cost is negligible at $p < 10^{-2}$ and grows honestly to 20–25% relative penalty only above $p = 10^{-2}$, consistent with the gate-level results of Section V-A.

## VI. SYNTHESIS COST ANALYSIS

| Resource | Standard CCX | Quantum-Adaptive KS($\varphi$) |
|---|---|---|
| Qubit count | 3 | 3 (no overhead) |
| T-gate count | 7 | 7+0 (QA-KS $\pi$) / 7+1 (QA-KS $\pi/2$)* |
| Macro-layers | 1 | 4 (strictly deeper) |
| Primitive depth | 5–6 | 5–6 + H, CP, H |
| Connectivity | Linear | Linear |
| Unitarity | Exact | Exact (verified) |
| Self-inverse | No | No |
| Error non-amplif. | No | Yes ($q_1$=0 subspace) |
| Phase kickback | None | $\sin(\varphi/2)$ |
| H-conj. sensitivity | None | X + Z both |

* CP(Z) is Clifford (0 T-gates); CP(S) costs 1 T-gate.

Table III. Resource Comparison — Standard CCX vs. Quantum-Adaptive KS($\varphi$).

## VII. OPEN PROBLEMS

**A.** Phase-channel error sensing characterization. Simulation confirmed that amplitude-channel error propagation is structurally identical between CCX and QA-KS($\varphi$) on $q_1$=0 basis inputs. The distinctive contribution of QA-KS($\varphi$) lies in the phase channel: the CP kickback imprints a coherent Z-type error indicator on $q_0$ not present in CCX. Formal characterization using phase-fidelity observables on superposition inputs, and a closed-form bound on phase-channel error sensitivity as a function of $\varphi$ and p, is required to fully substantiate this claim.

**B.** Pareto frontier at larger N. The coherent kickback chain confirms that QA-KS composition produces a distinct computation from CCX composition, with F=0.000 output fidelity on $q_0$=1 inputs. Circuit-level fidelity benchmark on adder extends this to standard algorithmic contexts. Scaling to N≥30 qubits via NVIDIA CUDA-Q GPU-accelerated simulation [21] is identified as the path to quantify kickback advantage at full code-distance scale.

**C.** Fault-tolerance pseudo-threshold. Formal threshold calculation [6] to determine whether error non-amplification yields net benefit over standard CCX at the logical qubit level.

**D.** Fredkin extension. A five-qubit extension with CSWAP-based palindrome and three-cell majority vote [17].

**E.** Hardware demonstration. Trapped-ion and neutral-atom platforms, where CP gates are native, are natural first targets for experimental validation.

**F**. Pre-syndrome integration with AI-accelerated decoders. The coherent phase signal imprinted on $q_0$ by the CP kickback gate constitutes a pre-syndrome error indicator that does not require classical readout. This is architecturally complementary to AI-accelerated decoding frameworks such as NVIDIA Ising [21], which operate on classical syndrome streams extracted from the quantum processor. A QA-KS(φ) gate embedded in a surface-code circuit may bias the local error distribution toward patterns more easily resolved by CNN-based pre-decoders, providing a primitive-level contribution to the decoding stack. Validation on NVIDIA CUDA-Q's GPU-accelerated noisy simulator would enable the circuit-level benchmarks outlined in Open Problem B at code distances beyond the reach of CPU-based statevector simulation.

## VIII. CONCLUSION

We have proposed and numerically verified the Quantum-Adaptive KS(φ) gate family. The name reflects parameter-conditioned behavior rooted in the QNCA majority-inspired bias rule, with Adaptive denoting compile-time φ-conditioning rather than runtime self-modification. Key contributions are: (1) the full 8 × 8 unitary matrix, confirmed exact and unitary to floating-point precision; (2) two canonical variants within the QA-KS(φ) family—QA-KS π/2 ($\varphi = \pi/2$, S gate) and QA-KS π ($\varphi = \pi$, Z gate)—that form a Pareto frontier over kickback amplitude, neither of which is self-inverse; (3) Qiskit noise simulation confirming honest fidelity cost: near-negligible at $p < 10^{-2}$; and (4) clarification of the gate's subspace structure—deterministic (up to phase) on $q_1 = 0$, producing four-component entangled superpositions on $q_1 = 1$.

These properties are achieved at 7 + 0 or 7 + 1 T-gates (QA-KS π and QA-KS π/2 respectively, versus 7 for standard CCX), with no qubit count increase, in a linear three-qubit connectivity layout directly compatible with existing surface-code and trapped-ion hardware.